\documentclass[preprint,prl,aps,amsmath,amssymb,color,superscriptaddress]{revtex4}
\usepackage{stmaryrd}
\usepackage{amsmath}
\usepackage{amssymb}
\usepackage{graphicx}
\usepackage{dcolumn}
\usepackage{bm}

\begin{document}

\begin{titlepage}

\title{Realizing intrinsic excitonic insulator by decoupling exciton binding energy from the minimum band gap}

\author{Zeyu Jiang}
\affiliation{State Key Laboratory of Low-Dimensional Quantum Physics and Collaborative Innovation Center of Quantum Matter, Department of Physics, Tsinghua University, Beijing 100084, China}

\author{Yuanchang Li}
\email{yuancli@bit.edu.cn}
\affiliation{Advanced Research Institute of Multidisciplinary Science, Beijing Institute of Technology, Beijing 100081, China}

\author{Shengbai Zhang}
\affiliation{Department of Physics, Applied Physics and Astronomy, Rensselaer Polytechnic Institute, Troy, NY, 12180, USA}

\author{Wenhui Duan}
\affiliation{State Key Laboratory of Low-Dimensional Quantum Physics and Collaborative Innovation Center of Quantum Matter, Department of Physics, Tsinghua University, Beijing 100084, China}
\affiliation{Institute for Advanced Study, Tsinghua University, Beijing 100084, China}

\date{\today}

\begin{abstract}
Direct-gap materials hold promises for excitonic insulator. In contrast to indirect-gap materials, here the difficulty to distinguish from a Peierls charge density wave is circumvented. However, direct-gap materials still suffer from the divergence of polarizability when the band gap approaches zero, leading to diminishing exciton binding energy. We propose that one can decouple the exciton binding energy from the band gap in materials where band-edge states have the same parity. First-principles calculations of two-dimensional GaAs and experimentally mechanically exfoliated single-layer TiS$_3$ lend solid supports to the new principle.
\end{abstract}

\maketitle

\draft

\vspace{2mm}

\end{titlepage}

Excitonic insulator (EI) is a new state of matter with a many-body ground state. It was named in 1967\cite{Kohn}, where the exciton binding energy ($E_b$) exceeds the bandgap ($E_g$), leading to the renormalization of the single-electron band structure in a semiconductor or a semimetal against the spontaneous formation of excitons. Because the exciton is made of two fermions, it obeys the bosonic statistics on the scale larger than the exciton radius, therefore allowing for
a Bose condensation. As a naturally-formed electron-hole condensate, EI behaves as a perfect insulator for both charge and heat transport\cite{Rontani}, despite that both electrons and holes are ideal carriers
for them. Hence, EI represents a highly promising and uncharted frontier in condensed matter physics, especially in the vicinity of a transition between the EI and non-EI phases. The search for EI has lasted a half century but
compelling experimental evidence is still lacking\cite{Kogar}. Although some evidence has been provided very recently in the quantum-well system\cite{Du}, an ideal EI would be to identify the material with $E_b > E_g$ naturally. The few materials proposed as possible candidates include 1\emph{T}-TiSe$_2$\cite{Cercellier,Cazzaniga,Kogar}, Ta$_2$NiSe$_5$\cite{Mor,Lu,Wakisaka},
TmSe$_{0.45}$Te$_{0.55}$\cite{Bucher,Bronold}, iron pnictides superconductor\cite{Brydon}, CaB$_6$\cite{Zhitomirsky, Bascones}, and carbon nanotube\cite{Varsano}. In the early days, much attention was paid to materials with
interacting electron-hole pockets located at different regions in the Brillouin zone for semiconductors with a small band gap or semimetals with a small band overlap to minimize the effect of screening \cite{Sherrington}.
Unfortunately, however, as schematically illustrated in Fig. 1(a), perceived formation of indirect excitons is always accompanied by a strong structural distortion such as a charge density wave due to the finite momentum transfer
\textbf{q}, which makes it difficult to determine whether the observed instability is originated from an excitonic effect or a band-type Jahn-Teller distortion\cite{Kogar,Rossnagel}. For this reason, recent interests both in
theory\cite{Kaneko} and experiment\cite{Mor,Lu} have shifted to direct gap semiconductors such as Ta$_2$NiSe$_5$ where structural distortion can be quenched, in spite of its generally larger screening due to band edge transitions,
as shown in Fig. 1(b).

To realize an intrinsic EI, one can engineer the band structure, e.g., through an external field modulation, to increase $E_b$ and/or decrease $E_g$ such that $E_b > E_g$ in an otherwise trivial semiconductor or semimetal. Intuitively, it seems trivial since external means can always yield $E_g\rightarrow 0$, and then $E_b > E_g$ is straightforward. However, this is not the case because the $E_b$ and $E_g$ are closely correlated and $E_g\rightarrow 0$ generally leads to diminishing $E_b$. The reason is that the $E_b$ is determined by the system screening which is characterized by the polarizability $\varepsilon$. Within the random phase approximation and not considering the local field effects, the polarizability may be expressed\cite{usPRL, ZRLiu} as
\begin{equation}\label{(1)}
\varepsilon = A\sum_{c,v} \int_{\rm \textbf{k}}\frac{|\langle \emph{u}_{\emph{c}, \rm \textbf{k}} |\nabla_{\rm \textbf{k}}| \emph{u}_{\emph{v}, \rm \textbf{k}}\rangle|^2}{\emph{E}_{\emph{c}, \rm \textbf{k}} - \emph{E}_{\emph{v}, \rm \textbf{k}}}d\rm \textbf{k}.
\end{equation}
where $\emph{u}_{\emph{c}, \rm \textbf{k}}$ and $\emph{u}_{\emph{v}, \rm \textbf{k}}$ refer to the periodical parts of conduction and valence band Bloch states, respectively, and \textbf{k} is integrated over the first Brillouin zone. $A$ is a dimension-related coefficient. On the appearance, Eq. (1) exhibits an inverse relationship between $E_g$ and $\varepsilon$, as by definition $E_g$ is the smallest $\emph{E}_{\emph{c}, \rm \textbf{k}} - \emph{E}_{\emph{v}, \rm \textbf{k}}$ in a direct gap material, whereby contributing the most to $\varepsilon$. More importantly, Eq. (1) reveals that when $E_g$ approaches zero, $\varepsilon$ is going to diverge, leading to a negligible $E_b$.

So it becomes clear that, the two seemingly intimately-related physical quantities $E_g$ and $E_b$ have to be decoupled in order to alter them individually via external means. According to Eq. (1), this requires $|\langle \emph{u}_{\emph{c}, \rm \textbf{k}} |\nabla_{\rm \textbf{k}}| \emph{u}_{\emph{v}, \rm \textbf{k}}\rangle|$ = 0, corresponding to band-edge transitions so that $\varepsilon$ can be finite when $E_g\rightarrow 0$. In this way, the $E_b$ could have no response to the reduction of $E_g$, therefore allowing for $E_b > E_g$ via band engineering. Note that prevalently used quantum-well structures to investigate the exciton condensate are also within such a notion but utilizing the spatial separation of electron and hole to suppress the band-edge transitions. To this end, two-dimensional (2D) materials provide us a new opportunity for realizing the intrinsic EI, not only because of the orders-of-magnitude enhanced $E_b$ \cite{usPRL}, but also because the electronic properties can be more effectively controlled by applying an electric field or a strain\cite{Fiori}. This can be contrasted to three-dimensional materials for which tuning $E_b$ and/or $E_g$ over a wide range still represents a formidable task.

%fig01
\begin{figure}[tbp]
\includegraphics[width=0.8\columnwidth]{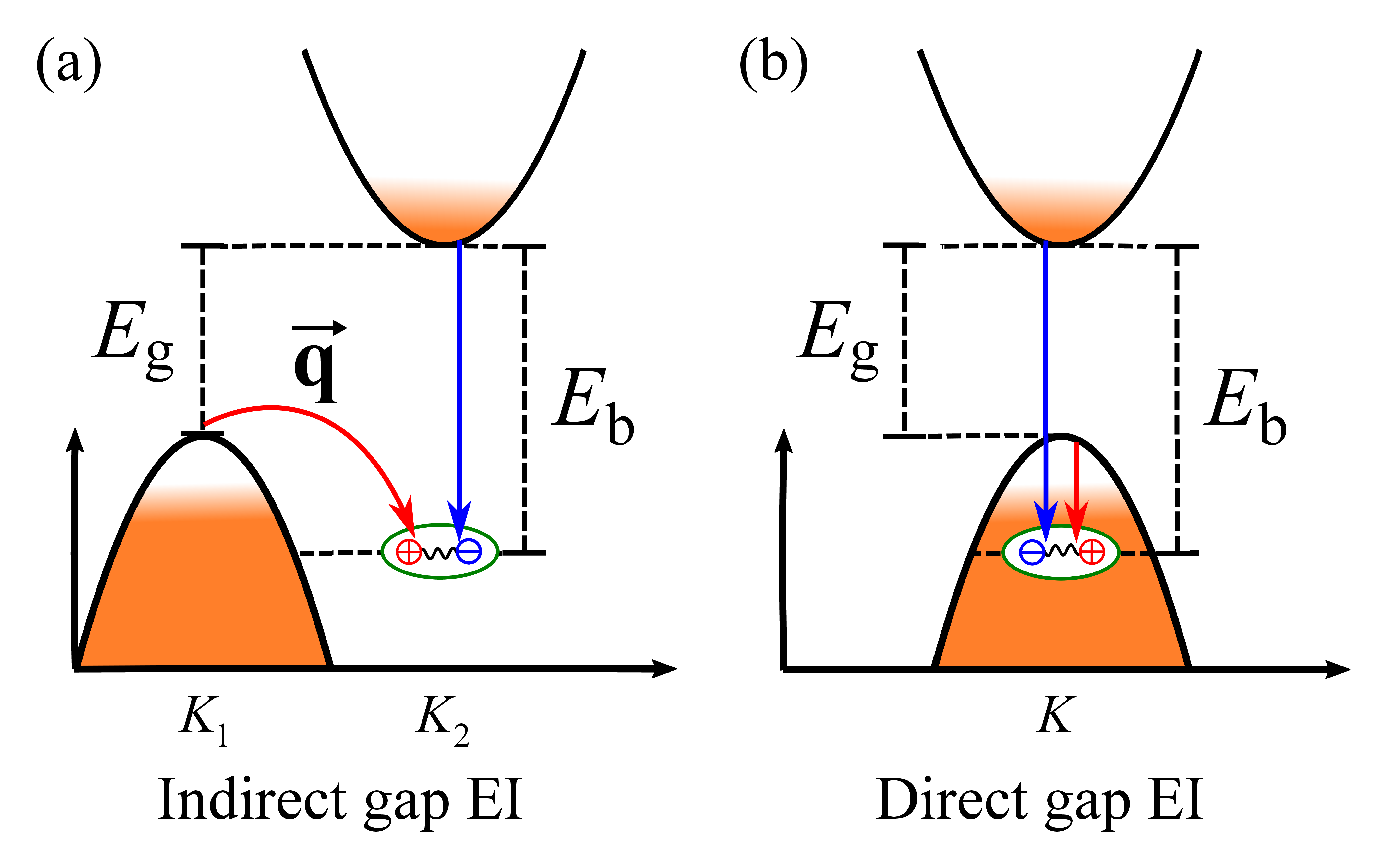}
\caption{\label{fig:fig1} (Color online) A schematic illustration of excitonic instability in (a) indirect- and (b) direct-gap materials. Typically, the former has a smaller dielectric screening but a larger tendency for structural distortion. The symbols ``$\oplus$" and ``$\ominus$" denote holes and electrons at the band edges, respectively. They form excitons through mutual Coulomb attraction, and the lower position of exciton states with respect to band-edge indicates the instability in energy of single-particle band structure against exciton formation.}
\end{figure}

In this work, we makes use of an intrinsic way to suppress band-edge transitions, namely, the parity, unlike the spatial separation in the quantum-well structures. That is, when the band-edge states have the same parity, transitions between them are dipole forbidden\cite{Nie} so $|\langle \emph{u}_{\emph{c}, \rm \textbf{k}} |\nabla_{\rm \textbf{k}}| \emph{u}_{\emph{v}, \rm \textbf{k}}\rangle|$ becomes very close to zero. As a result, the strongly related behavior\cite{usPRL} between $E_g$ and $E_b$ no longer holds, because the two are now derived from different states with different characteristic energies. In particular, $E_g$ is controlled, as usual, by the band-edge states, but $E_b$ is now controlled to a much lesser degree by such band-edge states but to a much larger degree by states away from the band edges. Consequently, the divergence of 2D polarizability as $E_g\rightarrow 0$ is prevented. In the following, we will first take the recently-proposed 2D GaAs\cite{Lucking} as a concrete example to illustrate how the principle come into play to result in the stabilization of the EI phase over the non-EI phase. Then we turn to the case of the mechanically exfoliated single-layer TiS$_3$ which would transit to an EI under a compressive strain about 3\%.

%fig02
\begin{figure}[tbp]
\includegraphics[width=0.8\columnwidth]{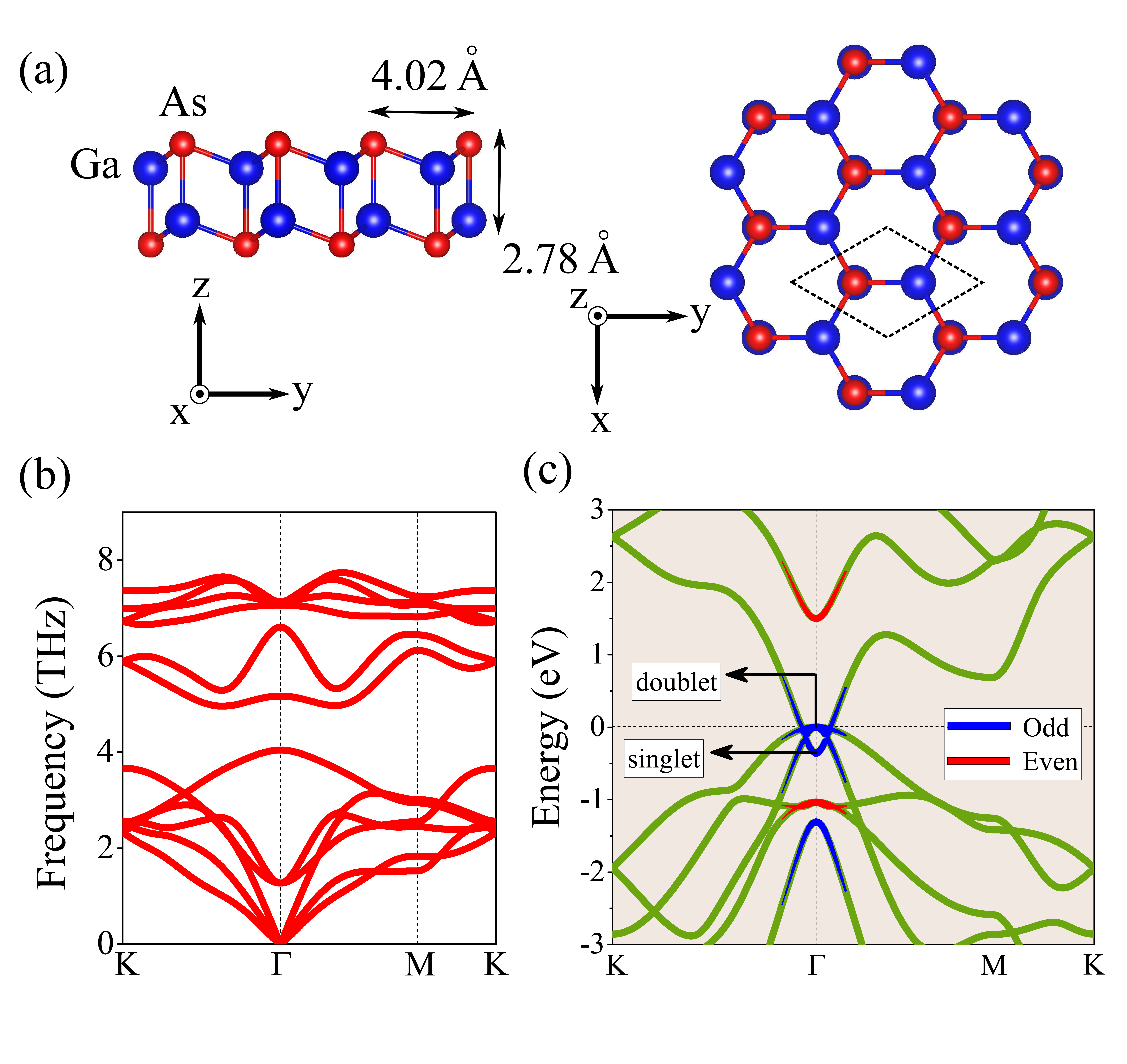}
\caption{\label{fig:fig2} (Color online) (a) Side (left) and top (right) views of 2D GaAs in the DLHC structure. Blue (large) and red (small) balls denote Ga and As atoms, respectively. Dashed rhombus denotes the unit cell. (b) The corresponding phonon spectrum. (c) PBE band structure of a 2D GaAs DLHC with band parities marked in color.}
\end{figure}

The density functional theory calculations were performed within the Perdew-Burke-Ernzerhof (PBE)\cite{PBE} exchange correlation functional as implemented in the VASP\cite{vasp} code. The plane-wave basis cutoff energy was set to 600 eV. An 18 \AA\ vacuum layer was used to avoid spurious interactions between adjacent layers. An $18 \times 18 \times 1$ $\Gamma$-centered $k$ grid was used to sample the Brillouin zone. The atomic structures were fully relaxed until residual forces on each atom were less than 0.001 eV/\AA\ and the system dynamical stability was further confirmed by phonon calculations. Owing to the well-known band-gap underestimation by PBE, we also performed Heyd-Scuseria-Ernzerhof (HSE) hybrid functional\cite{HSE03,HSE06} and many-body GW\cite{GWKress} calculations, and the results are given for comparison. We used Yambo\cite{yambo} code to calculate $E_b$ by solving the Bethe-Salpeter equation (BSE)\cite{BSELouie} with the single-electron band structure produced by Quantum Espresso package\cite{pwscf}.

Recently, a 2D form of traditional semiconductors has been synthesized via a migration-enhanced encapsulated growth technique utilizing epitaxial graphene\cite{Balushi}. In the meantime, based on the first-principles calculations, it was predicted that the ultra-thin limit of traditional binary III-V, II-VI, and I-VII semiconductors could take the kinetically stable and energetically favorable double-layer honeycomb (DLHC) structure\cite{Lucking}. Intriguingly, the DLHCs have the desired properties that band-edge states have the same symmetry. Figure 2(a) shows the DLHC structure for 2D GaAs. It is made of two monolayers of buckled honeycombs vertically coupled to each other with an AB stacking. Figure 2(b) shows the calculated phonon spectrum confirming its kinetic stability. In Ref. \onlinecite{Lucking}, ab initio molecular dynamics were also carried out to confirm the stability.

Figure 2(c) shows the PBE band structure for the GaAs. Near the Fermi energy, there are three bands, a singlet and the doubly degenerate bands (doublet) at $\Gamma$ point, which deserve special attention. Noticeably, these band edge states do have the same parity as required. However, 2D GaAs exhibits a metallic behavior instead of the usual semiconducting behavior and the doublet is above the singlet, which leads to a negative $E_g$ of $-$0.34 eV at $\Gamma$. While having the correct parities near the Fermi level, the metallic behavior is indicative of a strong screening, which usually diminishes $E_b$.

%fig03
\begin{figure}[tbp]
\includegraphics[width=0.75\columnwidth]{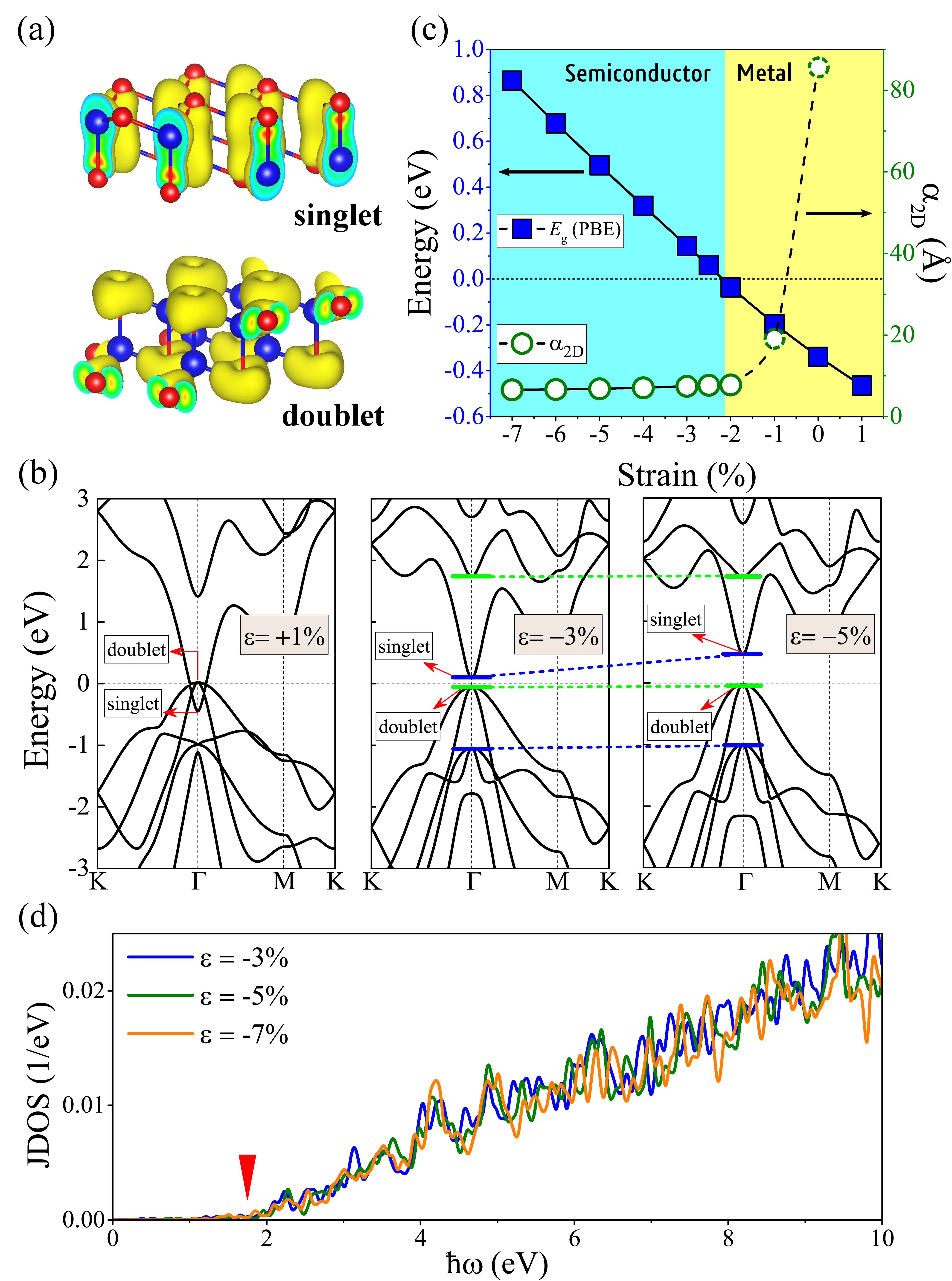}
\caption{\label{fig:fig3} (Color online) (a) Decomposed charge densities at the $\Gamma$ point for the singlet (upper panel) and doublet (lower panel) states, with an isosurface of $1.35 \times 10^{-4}$ e/\AA$^3$. (b) Band structure of the 2D GaAs under typical strains ($\varepsilon$). The two lowest-energy allowed transitions (between states of opposite parities) are marked in short green and blue horizontal bars, respectively. Fermi level is the energy zero. (c) Strain dependence of $E_g$ and corresponding $\alpha_{2D}$ at the PBE level. All the data points are obtained at the same calculation level for comparison, but note that $\alpha_{2D}$ eventually diverges in metal phase and the corresponding values may not be fully ``converged". (d) The JDOS under typical strains corresponding to a semiconducting 2D GaAs. Red arrow denotes the excitation energy after which the JDOS becomes significant.}
\end{figure}

To this end, we note that strain can induce metal-semiconductor transition in 2D materials\cite{Fiori,my2D}. Moreover, we find that the charge densities of the singlet and doublet at $\Gamma$ point have different out-of-plane and in-plane characters, as can be seen in Fig. 3(a). As a result, they must have substantially different responses to applied strain. Figure 3(b) plots selectively the band structures for 2D GaAs as a function of an in-plane biaxial strain. It can be seen that when the system is compressed by 3\%, band ``inversion" at $\Gamma$ point is lifted to open a gap of 0.14 eV. It increases notably with the strain to 0.49 eV at $-$5\% compression. In contrast, the 2D GaAs remains to be metallic under a tensile stain.

In Fig. 3(c), we plot the respective dependence of $E_g$ and 2D polarizability $\alpha_{2D}$ [derived from Eq. (1) with coefficient $A=\frac{e^{2}}{2\pi^{2}}$] on the in-plane biaxial strain. Clearly, they behave in completely different manners. While the $E_g$ reveals a simple linear dependence on the strain, the $\alpha_{2D}$ keeps almost unchanged for positive $E_g$ and rapidly diverges when the system becomes metallic. This strongly implies the quite different responses of $E_g$ and $E_b$ to strain, hence their decoupling, as will be quantitatively demonstrated later.	

We further plot the joint density of states (JDOS) under the typical strains in Fig. 3(d) in order to understand the nearly strain independent behavior of $\alpha_{2D}$. The JDOS is calculated as
\begin{equation}\label{(2)}
{\rm JDOS}(\omega) = \frac{\rm S}{2\pi^2} \sum_{c,v} \int_{\rm \textbf{k}} \delta(\emph{E}_{\emph{c}, \rm \textbf{k}} - \emph{E}_{\emph{v}, \rm \textbf{k}} - \hbar\omega) d^2\rm \textbf{k},
\end{equation}
where $S$ is the surface area of unit cell and the $\hbar\omega$ measures the excitation energy. Such a quantity characterizes the number of transitions between a certain energy range $\emph{E}_{\emph{c}, \rm \textbf{k}} - \emph{E}_{\emph{v}, \rm \textbf{k}}$. It is seen that the JDOS keeps negligible until an $\hbar\omega$ about 1.8 eV [Red arrow in Fig. 3(d)] and its distribution is almost invariant to the strain during the whole energy region, corresponding to the strain independence of $\alpha_{2D}$ as shown in Fig. 3(c). In addition, the existence of such a critical energy seems as if it was a strain-independent ``effective" $E_g$ of 1.8 eV that contributes to the system screening, although the system possesses a strain-sensitive electronic $E_g$ much smaller than that.

It is well-known that PBE underestimates $E_g$. In some cases, it can even be qualitatively wrong, e.g., predicting a semiconductor as a metal\cite{Tran}. Such a shortcoming can often be removed by using the HSE hybrid functional where a screened Coulomb potential is used for the Hartree-Fock exchange\cite{HSE03,HSE06}. Figure 4(a) depicts the HSE results as a function of the in-plane biaxial strain. It shows that 2D GaAs is a semiconductor with a gap of 0.25 eV, which transforms into a metal under a tensile strain of about 1\%. Figure 4(a) also shows $E_b$, calculated by the BSE approach\cite{BSELouie} at the HSE level. Stimulatingly, we see that the 2D GaAs is an intrinsic EI with $E_b$ exceeding $E_g$ in the strain range of $-2$\% to 1\%.

Figure 4(a) also sheds lights on the strain dependence of $E_g$ and $E_b$. While both $E_g$ and $E_b$ are linear functions of the strain, only $E_g$ is sensitive to the strain with a slope of $-0.2$ eV per 1\%-strain increase. In contrast, $E_b$ is nearly insensitive to the strain with a negligible slope of only $-8$ meV per 1\%-strain, until the system becomes metallic (not shown). So the ratio between the two is almost a factor of 25. Such a marked difference reinforces the notion that one can indeed decouple $E_b$ from $E_g$. One can qualitatively understand these results as follows: the band-edge states of 2D GaAs have different out-of-plane and in-plane characteristics for the singlet and doublet, which translate into the response of $E_g$ to strain. In contrast, $E_b$ (intrinsically the dielectric screening) is controlled by the overall effect of allowed transitions between the occupied and empty states according to Eq. (1). Figure 3(b) (middle and right panels) shows that not only the energy differences ($\emph{E}_{\emph{c}, \rm \textbf{k}} - \emph{E}_{\emph{v}, \rm \textbf{k}}$)  in this case are much larger than the minimum band gap ($E_g$), but also for both transitions (green $\rightarrow$ green and blue $\rightarrow$ blue), a non-band-edge state with a different strain response from those of the band-edge states is always involved. Not surprisingly, $E_b$ is no longer tied to $E_g$. Actually, our aforementioned results imply that they manifest themselves from the respectively strain-independent ``effective screening" gap and strain-sensitive electronic gap. Moreover, we plot the spatial distributions of the corresponding exciton state for 4\%-, 2.5\%-compressed and pristine GaAs, which represent the cases within the traditional semiconductor phase, near the phase boundary and within the exotic EI phase. No noticeable distinction is observed in both their shape and radius. Little change of the exciton state again corroborates the insensitivity of $E_b$ to strain.

%fig04
\begin{figure}[tbp]
\includegraphics[width=0.5\columnwidth]{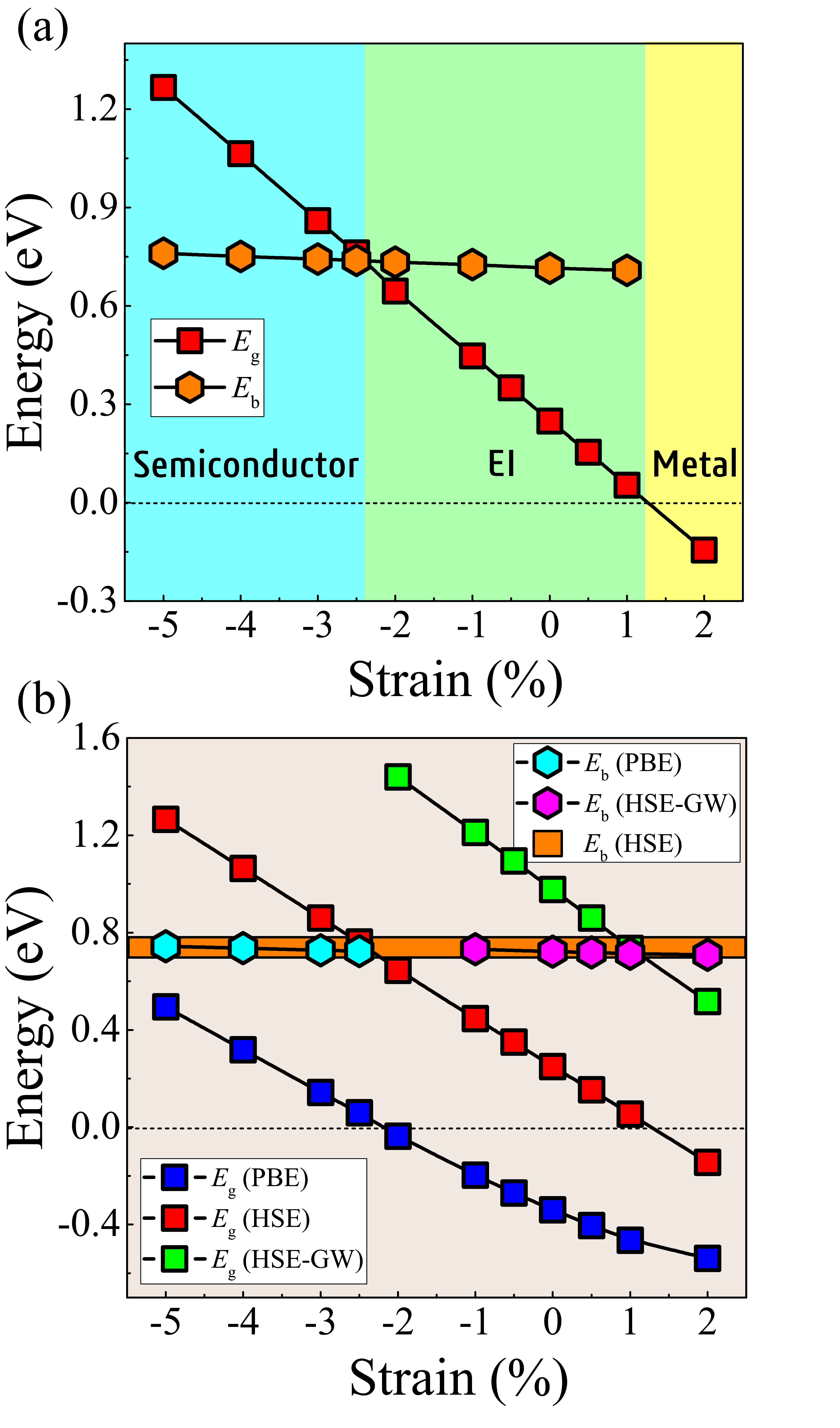}
\caption{\label{fig:fig4}(Color online) (a) Strain dependence of $E_g$ and $E_b$, calculated by HSE. It is evident that due to the decoupling between $E_g$ and $E_b$, phase transition from semiconducting to EI, and then to metallic phase takes place with increasing strain. (b) Same as in panel (a) but with different calculation methods, PBE, HSE, and GW (based on the HSE results). As a guide for the eye, $E_b$ by HSE is shown as a thick orange line.}
\end{figure}

In order to be certain of our findings, we also carry out many-body GW\cite{GWKress} calculations in a single-shot scheme ($\rm G_0 W_0$). The results are shown in Fig. 4(b). While unlike the HSE where the system is an EI without any strain, GW increases the $E_g$ from 0.25 eV (HSE) to 0.98 eV so the 2D GaAs under normal condition becomes a trivial band insulator, but turns into an EI at a modest tensile strain of 1\%. Three important points are worth noting: (1) as long as $E_g \ge 0$, $E_b$ is almost a constant and its value of about 0.73 eV is also insensitive to the calculation method. (2) The $E_g$ given by the different methods have a rather similar slope with respect to the strain. As we increase the level of accuracy of the calculation methods, $E_g$ exhibits a blue shift from PBE to HSE, and then to GW. (3) Irrespective of the methods, there is always a crossing point between the $E_g$ and $E_b$ curves in Fig. 4(b). Hence, irrespective of the technical details, we conclude that while the exact $E_g$ is difficult to predict, it is unambiguous that all the methods used here predicts the formation of EI at modest experimental conditions.

With these results in hand, we further notice that single-layer TiS$_3$, which has been experimentally exfoliated\cite{Island}, also fulfills the parity requirement. Previous work\cite{Dai} showed that the HSE calculation is necessary to yield the $E_g$ consistent with the experiment for TiS$_3$. Nevertheless, the computational cost is unaffordable at present for a fully converged solution of BSE at the HSE level for the system. Fortunately, it is revealed in Fig. 4(b) that the $E_b$ just weakly depends upon the calculation methods which suggests an alternative estimation of $E_b$ from the PBE result. Our first-principles calculations show that the $E_g$ (HSE level) monotonously decreases but the $E_b$ (PBE level) varies a little with the increase of compressive strain for the single-layer TiS$_3$. Without strain, it is 1.16 eV \emph{vs.} 0.92 eV for $E_g$ \emph{vs.} $E_b$, while it becomes 0.90 eV \emph{vs.} 0.94 eV under -3\% strain, indicative of the transition to EI phase. Such a moderate strain lies within an experimentally accessible regime, thus calling experimentalists for test.

In summary, we show that direct gap materials whose band-edge states possess the same parity are promising candidates for the EIs. Actually, any material with a lowest transition forbidden, regardless of direct or indirect gap, might be promising for engineering an intrinsic EI. Note that this EI principle works independent of the dimensionality. In three-dimensional bulk materials, however, the large screening often limits $E_b$ to be only a few or several tens of an meV, as well as making an effective tuning of $E_b$ and $E_g$ difficult. In this regard, 2D semiconductors with an appropriate band parity and a reasonable $E_g$ offer a unique opportunity for success. The 2D materials also hold another promise because a modest strain variation can lead to a rich phase diagram ranging from a traditional semiconductor, over an EI, to a metal, therefore potentially allowing for a device of complex functionalities to be made of purely a single material.

\begin{acknowledgments}
Work in China was supported by the Ministry of Science and Technology of China (Grant No. 2016YFA0301001), the National Natural Science Foundation of China (Grant Nos. 51788104, 11674071, 21373015, and 11674188), the Beijing Advanced Innovation
Center for Future Chip (ICFC), and Open Research Fund Program of the State Key Laboratory of Low-Dimensional Quantum Physics. Work in the US was supported by the US DOE Grant No. DESC0002623.
\end{acknowledgments}

\end{document}